# Evolution of an N-level system via automated vectorization of the Liouville equations and application to optically controlled polarization rotation

M.S. Shahriar[1,2], Ye Wang[1], S. Krishnamurthy[1], Y. Tu[1], G.S. Pati[3] and S. Tseng[1]

[1] Department of EECS, Northwestern University, Evanston, IL 60208, USA
[2] Department of Physics and Astronomy, Northwestern University, Evanston, IL 60208, USA
[3] Department of Physics & Pre-Engineering , Delaware State University, DE 19901, USA

E-mail: shahriar@northwestern.edu

**Abstract**
The Liouville equation governing the evolution of the density matrix for an atomic/molecular system is expressed in terms of a commutator between the density matrix and the Hamiltonian, along with terms that account for decay and redistribution. For finding solutions of this equation, it is convenient first to reformulate the Liouville equation by defining a vector corresponding to the elements of the density operator, and determining the corresponding time-evolution matrix. For a system of N energy levels, the size of the evolution matrix is $N^2 \times N^2$. When N is very large, evaluating the elements of these matrices becomes very cumbersome. We describe a novel algorithm that can produce the evolution matrix in an automated fashion for an arbitrary value of N. As a non-trivial example, we apply this algorithm to a fifteen-level atomic system used for producing optically controlled polarization rotation. We also point out how such a code can be extended for use in an atomic system with arbitrary number of energy levels.

## 1. Introduction

For some situations in atomic and molecular physics, it is necessary to consider a system with many energy levels, such as excitation involving many hyperfine levels and/or Zeeman sublevels. The Liouville equation that describes the evolution of the density matrix is expressed in terms of a commutator between the density matrix and the Hamiltonian, as well as additional terms that account for decay and redistribution [1-4]. To find solutions to this equation in steady-state or as a function of time, it is convenient first to reformulate the Liouville equation by defining a vector corresponding to the elements of the density operator, and determining the corresponding time evolution matrix. To find the steady-state solution in a closed system, it is also necessary to eliminate one of the diagonal elements of the density matrix from these equations, because of redundancy. For a system of N atoms, the size of the evolution matrix is $N^2 \times N^2$, and the size of the reduced matrix is $(N^2-1) \times (N^2-1)$. When N is very large, evaluating the elements of these matrices becomes very cumbersome. In this paper, we describe an algorithm that can produce the evolution matrix in an automated fashion, for an arbitrary value of N. We then apply this algorithm to a fifteen level atomic system used for producing optically controlled polarization rotation.

The paper is organized as follows. In section 2, we introduce the algorithm, using a two-level system as an example. In section 3, we verify the algorithm with a common three-level Raman system, and also show how to generate time independent Hamiltonian for any system by inspection alone. In section 4, we show how to generalize this to a system with arbitrary number of levels. In section 5, we use this algorithm to solve a 15-level atomic system used for producing optically controlled polarization rotation. In the Appendix, we include explicit Matlab codes for two-, three-, and fifteen-level systems and also a non-intuitive, but faster computational method for our algorithm.

## 2. A Two Level System

To illustrate the basic idea behind the algorithm, we first consider the simplest case: a two-level system of atoms excited by a monochromatic field [3], as illustrated in figure 1. Here, $\hbar\omega_1$ and $\hbar\omega_2$ are the energies of levels $|1\rangle$ and $|2\rangle$, and ω is the frequency of the laser, with a Rabi frequency of $\Omega_0$ [5].

The Hamiltonian, under electric dipole and rotating wave approximations, is given as

$$\mathcal{H}=\hbar\begin{pmatrix} \omega_1 & \frac{\Omega_0}{2}e^{i(\omega t-kz_0+\phi)} \\ \frac{\Omega_0}{2}e^{-i(\omega t-kz_0+\phi)} & \omega_2 \end{pmatrix} \qquad (1)$$

where k is the wavenumber of the laser, $z_0$ is the position of the atom, and ϕ is the phase of the field. Without loss of generality, we set $z_0=0$ and ϕ=0 in what follows. The corresponding two-level state vector for each atom is

$$|\psi\rangle = \begin{bmatrix} C_1(t) \\ C_2(t) \end{bmatrix}, \qquad (2)$$

which obeys the Schrodinger equation

$$i\hbar\frac{\partial|\psi\rangle}{\partial t} = \mathcal{H}|\psi\rangle \qquad (3)$$

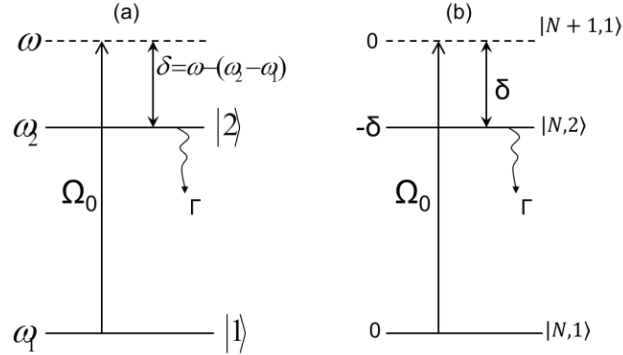

Figure 1: Schematic diagram showing a two-level system. (a): two-level system with eigenvector |1> and |2> ; (b): considering photon numbers, where |N,1> and |N+1,1> have the same energy, and the energy difference between |N+1,1> and |N,2> is $\hbar\delta$

To simplify the calculation, we convert the equations to the rotating wave frame by carrying out the following transformation into an interaction picture:

$$|\tilde{\psi}\rangle \equiv \begin{bmatrix} \tilde{C}_1(t) \\ \tilde{C}_2(t) \end{bmatrix} = R|\psi\rangle \tag{4a}$$

where

$$R = \begin{bmatrix} e^{i\omega_1 t} & 0 \\ 0 & e^{i\omega_2 t} \end{bmatrix} \tag{4b}$$

The Schroedinger equation now can be written as

$$i\hbar \frac{\partial |\tilde{\psi}\rangle}{\partial t} = \tilde{\mathcal{H}}|\tilde{\psi}\rangle \tag{5a}$$

where

$$\tilde{\mathcal{H}} = \hbar \begin{pmatrix} 0 & \frac{\Omega_0}{2} \\ \frac{\Omega_0}{2} & -\delta \end{pmatrix} \tag{5b}$$

$$\delta = \omega - (\omega_2 - \omega_1) \tag{5c}$$

The time independent Hamiltonian shown in equation 5b can also be derived easily without any algebraic manipulation. To see how, consider the diagram shown in figure 1(b), where we have added the number of photons as a quantum number in designating the quantum states. Thus, for example, $|N,1\rangle$ represents a joint quantum system where the number of photons in the laser field is N, and the atom is in state 1, and so on. Of course, a laser, being in a coherent state, is a linear superposition of number states, with a mean photon number $\langle N\rangle$, assumed to be much larger than unity. In the presence of such a field, the interaction takes place between near-degenerate states, namely $|N,2\rangle$ and $|N+1,1\rangle$, for example, with a coupling rate of $\Omega_0/2$, where $\Omega_N \propto \sqrt{N}$. Since the mean value of N is assumed to be very large, and much larger than its variance, one can assume the mean value of $\Omega_N$, defined as $\Omega_0$ to be proportional to $\sqrt{\langle N\rangle}$. Under this approximation, we see that the coupling between any neighboring, near-degenerate pair of states is $\Omega_0$, and the energies of these states differ by $\delta$. If we choose the energy of $|N+1,1\rangle$ to be 0, arbitrarily, then the energy of $|N,2\rangle$ is $-\hbar\delta$. The interaction is contained within a given manifold, so that a difference in energy (by $\hbar\omega$) between neighboring manifold is of no consequence in determining the evolution. These considerations directly lead to the Hamiltonian of equation 5(b). For a system involving more than two levels, a similar observation can be employed to write down the time-independent Hamiltonian by inspection, as we will show later.

The decay of the excited state amplitude, at the rate of $\Gamma/2$, can be taken into account by adding a complex term to the Hamiltonian, as follows:

$$\tilde{\mathcal{H}}' = \hbar \begin{bmatrix} 0 & \frac{\Omega_0}{2} \\ \frac{\Omega_0}{2} & -\frac{i\Gamma}{2} - \delta \end{bmatrix} \tag{6}$$

For this modified Hamiltonian, the equation of evolution for the interaction picture density operator can be expressed as

$$\frac{\partial}{\partial t}\tilde{\rho} = \frac{\partial}{\partial t}\tilde{\rho}_{ham} + \frac{\partial}{\partial t}\tilde{\rho}_{source} + \frac{\partial}{\partial t}\tilde{\rho}_{trans-decay} \equiv Q \tag{7}$$

where the 2nd term in the middle accounts for the influx of atoms into a state due to decay from another state, and the 3$^{rd}$ term stands for any dephasing unaccompanied by population decay, often called transverse decay. In the case of a two level system, we have:

$$\frac{\partial}{\partial t}\tilde{\rho}_{ham} = -\frac{i}{\hbar}[\tilde{\mathcal{H}}'\tilde{\rho} - \tilde{\rho}\tilde{\mathcal{H}}'^*] \tag{8a}$$

$$\frac{\partial}{\partial t}\tilde{\rho}_{source} = \begin{bmatrix} \Gamma\tilde{\rho}_{22} & 0 \\ 0 & 0 \end{bmatrix} \tag{8b}$$

$$\frac{\partial}{\partial t}\tilde{\rho}_{trans-decay} = \begin{bmatrix} 0 & -\gamma_d\tilde{\rho}_{12} \\ -\gamma_d\tilde{\rho}_{21} & 0 \end{bmatrix} \tag{8c}$$

For simplicity, we ignore the dephasing term in 8c.

Substituting eqn. 6 into eqn. 8a, we get:

$$\frac{\partial}{\partial t}\tilde{\rho}_{ham} = \begin{pmatrix} \frac{1}{2}i\Omega_0(\tilde{\rho}_{12}-\tilde{\rho}_{21}) & \frac{1}{2}i\left((i\Gamma-2\delta)\tilde{\rho}_{12}+\Omega_0(\tilde{\rho}_{11}-\tilde{\rho}_{22})\right) \\ -\frac{1}{2}i\left((-i\Gamma-2\delta)\tilde{\rho}_{21}+\Omega_0(\tilde{\rho}_{11}-\tilde{\rho}_{22})\right) & \frac{1}{2}\left(-i\Omega_0(\tilde{\rho}_{12}-\tilde{\rho}_{21})-2\Gamma\tilde{\rho}_{22}\right) \end{pmatrix}$$

(10)

Substituting eqns. 8 and 10 into eqn. 7, we get

$$\frac{\partial}{\partial t}\tilde{\rho} = \frac{\partial}{\partial t}\begin{pmatrix} \tilde{\rho}_{11} & \tilde{\rho}_{12} \\ \tilde{\rho}_{21} & \tilde{\rho}_{22} \end{pmatrix} = \begin{pmatrix} \frac{1}{2}i\Omega_0(\tilde{\rho}_{12}-\tilde{\rho}_{21})+\Gamma\tilde{\rho}_{22} & \frac{1}{2}i\left(i(\Gamma+2i\delta)\tilde{\rho}_{12}+\Omega_0(\tilde{\rho}_{11}-\tilde{\rho}_{22})\right) \\ -\frac{1}{2}i\left((-i\Gamma-2\delta)\tilde{\rho}_{21}+\Omega_0(\tilde{\rho}_{11}-\tilde{\rho}_{22})\right) & \frac{1}{2}\left(-i\Omega_0(\tilde{\rho}_{12}-\tilde{\rho}_{21})-2\Gamma\tilde{\rho}_{22}\right) \end{pmatrix}$$

$$= Q \equiv \begin{pmatrix} Q_{11} & Q_{12} \\ Q_{21} & Q_{22} \end{pmatrix}$$

(11)

In general, each of the matrix elements $Q_{ij}$ can depend on all the $\rho_{ij}$. In order to find the steady state solution, it is convenient to construct the following vector

$$A = \begin{bmatrix} \tilde{\rho}_{11} \\ \tilde{\rho}_{12} \\ \tilde{\rho}_{21} \\ \tilde{\rho}_{22} \end{bmatrix} \tag{12}$$

Equation 11 can now be expressed as a matrix equation

$$\frac{\partial}{\partial t}A = MA \tag{13}$$

where M is a (4×4) matrix, represented formally as:

$$M = \begin{bmatrix} M_{11} & M_{12} & M_{13} & M_{14} \\ M_{21} & M_{22} & M_{23} & M_{24} \\ M_{31} & M_{32} & M_{33} & M_{34} \\ M_{41} & M_{42} & M_{43} & M_{44} \end{bmatrix}$$

Of course, the elements of this matrix can be read-off from eqn. 11. However, this task is quite cumbersome for an N-level system. Thus, it is useful to seek a general rule for finding this element without having to write down eqn. 11 explicitly. Later on in this paper, we establish such a rule, and specify the algorithm for implementing it. Here, we can illustrate this rule with some explicit examples:

$M_{11}=Q_{11}$, if we set $\tilde{\rho}_{11}=1$ and $\tilde{\rho}_{ij(ij \neq 11)}=0$ in eqn. 7;

$M_{12}=Q_{11}$, if we set $\tilde{\rho}_{12}=1$ and $\tilde{\rho}_{ij(ij \neq 12)}=0$ in eqn. 7;

$M_{13}=Q_{11}$, if we set $\tilde{\rho}_{21}=1$ and $\tilde{\rho}_{ij(ij \neq 21)}=0$ in eqn. 7;

$M_{14}=Q_{11}$, if we set $\tilde{\rho}_{22}=1$ and $\tilde{\rho}_{ij(ij \neq 22)}=0$ in eqn. 7;

$M_{21}=Q_{12}$, if we set $\tilde{\rho}_{11}=1$ and $\tilde{\rho}_{ij(ij \neq 11)}=0$ in eqn. 7;

$M_{22}=Q_{12}$, if we set $\tilde{\rho}_{12}=1$ and $\tilde{\rho}_{ij(ij \neq 12)}=0$ in eqn. 7;

$M_{23}=Q_{12}$, if we set $\tilde{\rho}_{21}=1$ and $\tilde{\rho}_{ij(ij \neq 21)}=0$ in eqn. 7;

$M_{24}=Q_{12}$, if we set $\tilde{\rho}_{22}=1$ and $\tilde{\rho}_{ij(ij \neq 22)}=0$ in eqn. 7;

and so on... (14)

This is the key element of the algorithm presented in this paper. Explicitly, in a computer program, such as the one in Appendix A, every time a parameter is changed, the elements of the M matrix are obtained by evaluating equation 7, while setting all but one of the elements of the density matrix to zero. For numerical integration as a function of time, one can then use a Taylor expansion to solve equation 13.

To find the steady-state solution, we set $\frac{\partial}{\partial t} A = 0$, so that:

$$\begin{bmatrix} M_{11} & M_{12} & M_{13} & M_{14} \\ M_{21} & M_{22} & M_{23} & M_{24} \\ M_{31} & M_{32} & M_{33} & M_{34} \\ M_{41} & M_{42} & M_{43} & M_{44} \end{bmatrix} \begin{bmatrix} \tilde{\rho}_{11} \\ \tilde{\rho}_{12} \\ \tilde{\rho}_{21} \\ \tilde{\rho}_{22} \end{bmatrix} = 0 \quad (15)$$

Expanding this equation, we get:

$$\begin{cases} M_{11}\tilde{\rho}_{11} + M_{12}\tilde{\rho}_{12} + M_{13}\tilde{\rho}_{21} = -M_{14}\tilde{\rho}_{22} \\ M_{21}\tilde{\rho}_{11} + M_{22}\tilde{\rho}_{12} + M_{23}\tilde{\rho}_{21} = -M_{24}\tilde{\rho}_{22} \\ M_{31}\tilde{\rho}_{11} + M_{32}\tilde{\rho}_{12} + M_{33}\tilde{\rho}_{21} = -M_{34}\tilde{\rho}_{22} \\ M_{41}\tilde{\rho}_{11} + M_{42}\tilde{\rho}_{12} + M_{43}\tilde{\rho}_{21} = -M_{44}\tilde{\rho}_{22} \end{cases} \quad (16)$$

For a closed system, there cannot be any net influx or outflux of atoms from the system. Thus, the rate of change of one of the diagonal (population) terms of the density matrix is the negative sum of the rates of change of the other diagonal (population) terms. Thus, one of the equations in the above system of equations is rendered redundant. We also know that for a closed system, sum of the diagonal elements of the density matrix equals unity. In the case of the two level system, we thus have $\tilde{\rho}_{11} + \tilde{\rho}_{22} = 1$. We can thus choose to eliminate the last equation, for example, and replace $\tilde{\rho}_{22}$ with $(1 - \tilde{\rho}_{11})$ in the remaining three equations, to get

$$\begin{bmatrix} M_{11} & M_{12} & M_{13} \\ M_{21} & M_{22} & M_{23} \\ M_{31} & M_{32} & M_{33} \end{bmatrix} \begin{bmatrix} \tilde{\rho}_{11} \\ \tilde{\rho}_{12} \\ \tilde{\rho}_{21} \end{bmatrix} \equiv M' \begin{bmatrix} \tilde{\rho}_{11} \\ \tilde{\rho}_{12} \\ \tilde{\rho}_{21} \end{bmatrix} = \begin{bmatrix} M_{14} \\ M_{24} \\ M_{34} \end{bmatrix} \tilde{\rho}_{11} - \begin{bmatrix} M_{14} \\ M_{24} \\ M_{34} \end{bmatrix} \quad (17a)$$

so that

$$\begin{bmatrix} (M_{11} - M_{14}) & M_{12} & M_{13} \\ (M_{21} - M_{24}) & M_{22} & M_{23} \\ (M_{31} - M_{34}) & M_{32} & M_{33} \end{bmatrix} \begin{bmatrix} \tilde{\rho}_{11} \\ \tilde{\rho}_{12} \\ \tilde{\rho}_{21} \end{bmatrix} = - \begin{bmatrix} M_{14} \\ M_{24} \\ M_{34} \end{bmatrix} \quad (17b)$$

Here, we have defined M' as the reduced matrix resulting from M after eliminating the last row and column, for convenience of discussion during the presentation of the general algorithm later on. To simplify the notation further, we define:

$$B \equiv \begin{bmatrix} \tilde{\rho}_{11} \\ \tilde{\rho}_{12} \\ \tilde{\rho}_{21} \end{bmatrix}, \ S \equiv \begin{bmatrix} M_{14} \\ M_{24} \\ M_{34} \end{bmatrix}, \ W \equiv \begin{bmatrix} (M_{11} - M_{14}) & M_{12} & M_{13} \\ (M_{21} - M_{24}) & M_{22} & M_{23} \\ (M_{31} - M_{34}) & M_{32} & M_{33} \end{bmatrix} \quad (18)$$

Using these definitions in eqn. 17, we get:
$$WB = -S$$
Thus, the steady-state solution is simply given by:
$$B = -W^{-1}S \quad (19)$$

In a computer code, such as the one in Appendix A, the elements of W and S can be determined in an automated fashion by using a simple algorithm based on a generalization of this example. We get the values of $\tilde{\rho}_{11}$, $\tilde{\rho}_{12}$, and $\tilde{\rho}_{21}$ by using eqn. 19. Using the condition $\tilde{\rho}_{11} + \tilde{\rho}_{22} = 1$, we can then find the value of $\tilde{\rho}_{22}$.

For the 2-level system, the elements of M, W and S can be worked out by hand, without employing the general rules, with relative ease. However, for arbitrarily large systems, it can become exceedingly cumbersome. In what follows, we describe a compact algorithm for determining the elements of M, W and S for a system with N energy levels.

To start with, determine the elements of the complex effective Hamiltonian of eqn. 6, as well as the elements of $\tilde{\rho}_{source}$ for the N-level system. These matrices can be used to calculate the elements of Q, as defined in eqns. 7 and 11. The elements of M can then be found by using the following algorithm. Let $M_{np}$ denote the element corresponding to the n-th row and p-th column of the M matrix. Similarly, let $Q_{\alpha\beta}$ denote the element corresponding to the α-th row and β-th column of the Q matrix, and $\tilde{\rho}_{\varepsilon\sigma}$ denote the elements corresponding to the ε-th row and σ-th column of the $\tilde{\rho}$ matrix. Then one can use the following prescription to obtain $M_{np}$:

$M_{np} = Q_{\alpha\beta}$ if we set $\tilde{\rho}_{\varepsilon\sigma} = 1$ and $\tilde{\rho}_{ij(ij \neq \varepsilon\sigma)} = 0$ in eqn. 7.

Thus, the crux of the algorithm is to obtain a way of finding α,β,ε and γ efficiently, for a given set of values of {n,p}. These indices are obtained as follows:

$$\beta = \mathit{nzrem}[n/N]; \quad \alpha = 1+(n-\beta)/N; \quad \sigma = \mathit{nzrem}[p/N]; \quad \varepsilon = 1+(p-\sigma)/N \qquad (20)$$

where *nzrem* is a user-defined function prescribed as follows: *nzrem*[A/B]= *remainder*[A/B] if the remainder is non-zero; otherwise *nzrem*[A/B]=B. As an example, consider the case of the last line in eqn. 14. Here, n=2, p=4 and N=2. Thus, applying eqn.20, we get: β=2, α=1, σ=2, ε=2, in agreement with the last line of eqn. 14. We should note that there are other ways to determine these coefficients as well, using the *greatest integer* function, for example.

Once $(\alpha,\beta)$ and $(\varepsilon,\sigma)$ have been obtained, set $\tilde{\rho}_{\varepsilon\sigma}$ to be 1 while setting the other elements to 0, evaluate the Q matrix using eqn. 7, and then pick out $Q_{\alpha\beta}$ and assign it to $M_{np}$. Then repeat this procedure of evaluating the Q matrix every time with different element of the $\tilde{\rho}$ matrix set to 1 sequentially, until all elements of the M matrix have been calculated.

The steps for finding S and W, as defined in eqn. 18 for the case of a two level system, are rather simple. The last column of the M matrix barring the very last element is the S matrix. In order to determine the elements of W, find first the M' matrix, which is obatined from M by eliminating the last row and the last column, as illustrated in eqn. 17a for a two level system. Define $W_i$ and $M'_i$ as the *i*-th column of the W and the M' matrix. Then, update a selected set of $W_i$, using an index *k* running from 1 to (N-1), as follows:

$$W_{(k-1)N+k} = M'_{(k-1)N+k} - S \qquad (21)$$

To illustrate this rule, consider, for example, the case where N=3. In this case, $W_1=M_1-S$ (for k=1) and $W_5=M_5-S$ (for k=2), and the other six columns remain the same. With S and W thus determined, eqn. 20 is used to find the steady-state solution vector: B. A particular element of the density matrix, $\tilde{\rho}_{jk}$ (excluding $\tilde{\rho}_{NN}$), corresponds to the ((j-1)*N+k)-th element of the B vector. The population in the N-th level, $\tilde{\rho}_{NN}$ is simply obtained from the knowledge of the steady- state populations in all other levels and the constraint $\sum_{i=1}^{N}\tilde{\rho}_{ii} = 1$. Explicitly, we can write:

$$\tilde{\rho}_{NN} = 1 - \sum_{j=1}^{(N-1)} B((j-1)N+j) \qquad (22)$$

where we have used the notation that B(k) represents the k-th element of the B vector.

A Matlab code for an N-level system, applied to the case of two-levels, is shown in Appendix A. The code is valid for a general system, only N (number of levels in the system), and the effective, complex Hamiltonian (eqn.6) and the source terms (eqn. 8) need to be changed. *The rest of the program does not have to be changed.* Of course, the plotting commands would have to be defined by the user based on the information being sought. As an example, the population of the excited state as a function of the detuning, $\delta$, produced by this code, is plotted in figure 2.

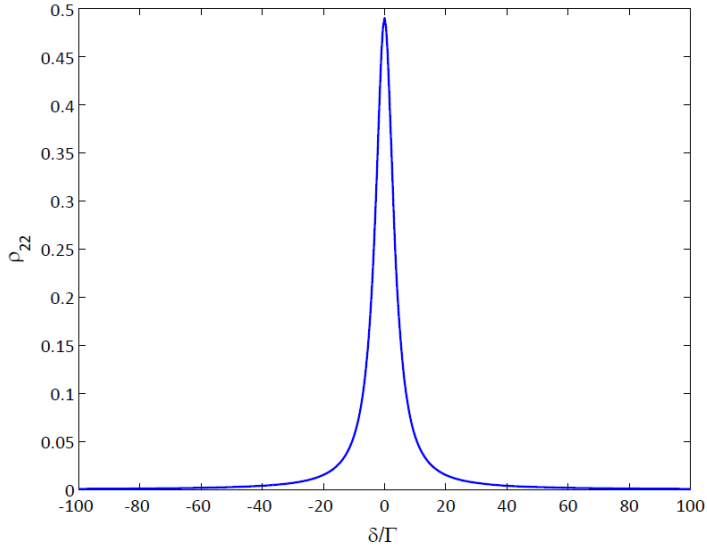

Figure 2. Population of excited state for a two-level system calculated using this algorithm.

## 3. A Three Level System

The two-level problem discussed above is somewhat trivial, and may mask the generality of the algorithm. Therefore, we include here the specific steps for a three-level $\Lambda$ system [6-11], shown in figure 3, in order to elucidate how the algorithm is completely scalable to an arbitrary number of energy levels involved.

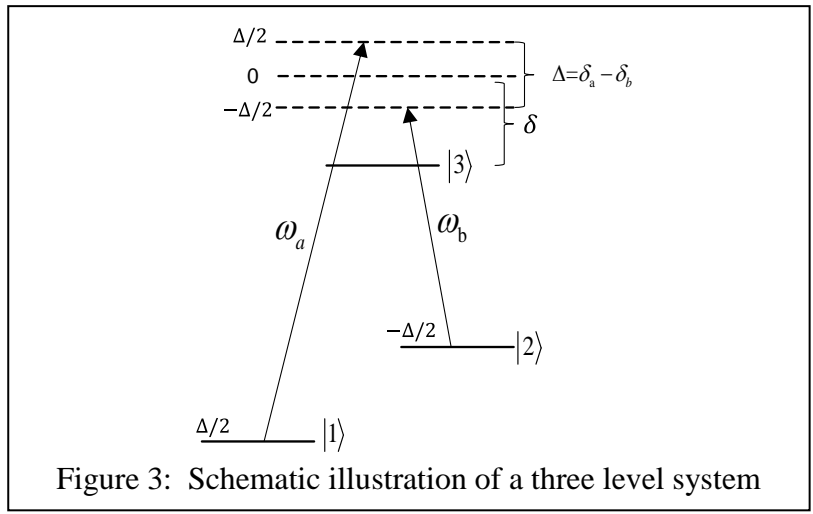

Figure 3: Schematic illustration of a three level system

In this case, the Hamiltonian under electric dipole and rotating wave approximations is given by

$$\mathcal{H}=\hbar\begin{pmatrix} \omega_1 & 0 & \frac{\Omega_a}{2}e^{i\omega_a t} \\ 0 & \omega_2 & \frac{\Omega_b}{2}e^{i\omega_b t} \\ \frac{\Omega_a}{2}e^{i\omega_a t} & \frac{\Omega_b}{2}e^{i\omega_b t} & \omega_3 \end{pmatrix} \quad (23)$$

where $\hbar\omega_1$, $\hbar\omega_2$, and $\hbar\omega_3$ are the energies of the three levels, and $\omega_a$ and $\omega_b$ are the frequencies of the laser fields.

After applying the interaction picture transformation using the following matrix

$$R=\begin{bmatrix} e^{i\theta t} & 0 & 0 \\ 0 & e^{i\beta t} & 0 \\ 0 & 0 & e^{i\varsigma t} \end{bmatrix} \quad (24)$$

where $\theta = \omega_1 - \frac{\Delta}{2}, \beta = \omega_2 + \frac{\Delta}{2}, \Delta = \delta_a - \delta_b, \delta = (\delta_a + \delta_b)/2, \delta_a = \omega_a - (\omega_3 - \omega_1), \delta_b = \omega_b - (\omega_3 - \omega_2)$ the Hamiltonian can be expressed as

$$\tilde{\mathcal{H}}=\frac{\hbar}{2}\begin{pmatrix} \Delta & 0 & \Omega_a \\ 0 & -\Delta & \Omega_b \\ \Omega_a & \Omega_b & -2\delta \end{pmatrix} \quad (25)$$

The transformed state vector for each atom can be written as

$$|\tilde{\psi}\rangle = R|\psi\rangle = \begin{bmatrix} \tilde{C}_1(t) \\ \tilde{C}_2(t) \\ \tilde{C}_3(t) \end{bmatrix} \quad (26)$$

The time independent Hamiltonian $\tilde{H}$ of equation 25 can be written down by inspection, following the discussion presented earlier for the two-level system. First, we observe that the energy difference between |1> and |3> ($\tilde{H}_{11} - \tilde{H}_{33}$) is $\hbar\delta_a$, and the energy difference between |2> and |3> ($\tilde{H}_{22} - \tilde{H}_{33}$) is $\hbar\delta_b$. Next, we make a judicious but arbitrary choice that $\tilde{H}_{11} = \frac{\hbar}{2}\Delta$. We then get that $\tilde{H}_{33} = -\hbar\delta$, which in turn implies that $\tilde{H}_{22} = -\frac{\hbar}{2}\Delta$. The off diagonal terms are, of course, obvious, with non-zero elements for transitions excited by fields. This approach is generic, and can be used to find the time independent Hamiltonian by inspection for an arbitrary number of levels. We should note that a complication exists when closed-loop excitations are present. In that case, it is wiser to work out the Hamiltonian explicitly using the transformation matrix approach outlined here. We now add the decay term to get the complex Hamiltonian

$$\tilde{\mathcal{H}}' = \frac{\hbar}{2} \begin{pmatrix} \Delta & 0 & \Omega_a \\ 0 & -\Delta & \Omega_b \\ \Omega_a & \Omega_b & -i\Gamma - 2\delta \end{pmatrix} \tag{27}$$

We assume that the population of the excited state decays at the same rate ($\Gamma/2$) from $|3\rangle$ to $|1\rangle$ and from $|3\rangle$ to $|2\rangle$. Now we construct the M matrix for the three-level system which satisfies the following equation under the steady-state condition:

$$\begin{bmatrix} M_{11} & M_{12} & M_{13} & M_{14} & M_{15} & M_{16} & M_{17} & M_{18} & M_{19} \\ M_{21} & M_{22} & M_{23} & M_{24} & M_{25} & M_{26} & M_{27} & M_{28} & M_{29} \\ M_{31} & M_{32} & M_{33} & M_{34} & M_{35} & M_{36} & M_{37} & M_{38} & M_{39} \\ M_{41} & M_{42} & M_{43} & M_{44} & M_{45} & M_{46} & M_{47} & M_{48} & M_{49} \\ M_{51} & M_{52} & M_{53} & M_{54} & M_{55} & M_{56} & M_{57} & M_{58} & M_{59} \\ M_{61} & M_{62} & M_{63} & M_{64} & M_{65} & M_{66} & M_{67} & M_{68} & M_{69} \\ M_{71} & M_{72} & M_{73} & M_{74} & M_{75} & M_{76} & M_{77} & M_{78} & M_{79} \\ M_{81} & M_{82} & M_{83} & M_{84} & M_{85} & M_{86} & M_{87} & M_{88} & M_{89} \\ M_{91} & M_{92} & M_{93} & M_{94} & M_{95} & M_{96} & M_{97} & M_{98} & M_{99} \end{bmatrix} \begin{bmatrix} \tilde{\rho}_{11} \\ \tilde{\rho}_{12} \\ \tilde{\rho}_{13} \\ \tilde{\rho}_{21} \\ \tilde{\rho}_{22} \\ \tilde{\rho}_{23} \\ \tilde{\rho}_{31} \\ \tilde{\rho}_{32} \\ \tilde{\rho}_{33} \end{bmatrix} = 0 \tag{28}$$

The elements of the M matrix can be found explicitly by following the same steps as shown in equation (7) through (13) for the two-level system. Alternatively, these can be found by using the algorithmic approach outlined in equation (20), and implemented by a computer code. The M-matrix can be obtained in O ($N^2$) steps as opposed O ($N^4$) that would be needed using the method prescribed thus far, but it is non-intuitive and masks the understanding of the algorithm. We have outlined the faster method in the appendix.

Substituting $\tilde{\rho}_{11} + \tilde{\rho}_{22} + \tilde{\rho}_{33} = 1$ into eqn. 28, we get

$$\begin{bmatrix} M_{11} & M_{12} & M_{13} & M_{14} & M_{15} & M_{16} & M_{17} & M_{18} \\ M_{21} & M_{22} & M_{23} & M_{24} & M_{25} & M_{26} & M_{27} & M_{28} \\ M_{31} & M_{32} & M_{33} & M_{34} & M_{35} & M_{36} & M_{37} & M_{38} \\ M_{41} & M_{42} & M_{43} & M_{44} & M_{45} & M_{46} & M_{47} & M_{48} \\ M_{51} & M_{52} & M_{53} & M_{54} & M_{55} & M_{56} & M_{57} & M_{58} \\ M_{61} & M_{62} & M_{63} & M_{64} & M_{65} & M_{66} & M_{67} & M_{68} \\ M_{71} & M_{72} & M_{73} & M_{74} & M_{75} & M_{76} & M_{77} & M_{78} \\ M_{81} & M_{82} & M_{83} & M_{84} & M_{85} & M_{86} & M_{87} & M_{88} \end{bmatrix} \begin{bmatrix} \tilde{\rho}_{11} \\ \tilde{\rho}_{12} \\ \tilde{\rho}_{13} \\ \tilde{\rho}_{21} \\ \tilde{\rho}_{22} \\ \tilde{\rho}_{23} \\ \tilde{\rho}_{31} \\ \tilde{\rho}_{32} \end{bmatrix} = \begin{bmatrix} M_{19} \\ M_{29} \\ M_{39} \\ M_{49} \\ M_{59} \\ M_{69} \\ M_{79} \\ M_{89} \end{bmatrix} \tilde{\rho}_{11} + \begin{bmatrix} M_{19} \\ M_{29} \\ M_{39} \\ M_{49} \\ M_{59} \\ M_{69} \\ M_{79} \\ M_{89} \end{bmatrix} \tilde{\rho}_{22} - \begin{bmatrix} M_{19} \\ M_{29} \\ M_{39} \\ M_{49} \\ M_{59} \\ M_{69} \\ M_{79} \\ M_{89} \end{bmatrix}$$

(29.a)

Or

$$\begin{bmatrix} (M_{11}-M_{19}) & M_{12} & M_{13} & M_{14} & (M_{15}-M_{19}) & M_{16} & M_{17} & M_{18} \\ (M_{21}-M_{29}) & M_{22} & M_{23} & M_{24} & (M_{25}-M_{29}) & M_{26} & M_{27} & M_{28} \\ (M_{31}-M_{39}) & M_{32} & M_{33} & M_{34} & (M_{35}-M_{39}) & M_{36} & M_{37} & M_{38} \\ (M_{41}-M_{49}) & M_{42} & M_{43} & M_{44} & (M_{45}-M_{49}) & M_{46} & M_{47} & M_{48} \\ (M_{51}-M_{59}) & M_{52} & M_{53} & M_{54} & (M_{55}-M_{59}) & M_{56} & M_{57} & M_{58} \\ (M_{61}-M_{69}) & M_{62} & M_{63} & M_{64} & (M_{65}-M_{69}) & M_{66} & M_{67} & M_{68} \\ (M_{71}-M_{79}) & M_{72} & M_{73} & M_{74} & (M_{75}-M_{79}) & M_{76} & M_{77} & M_{78} \\ (M_{81}-M_{89}) & M_{82} & M_{83} & M_{84} & (M_{85}-M_{89}) & M_{86} & M_{87} & M_{88} \end{bmatrix} \begin{bmatrix} \tilde{\rho}_{11} \\ \tilde{\rho}_{12} \\ \tilde{\rho}_{13} \\ \tilde{\rho}_{21} \\ \tilde{\rho}_{22} \\ \tilde{\rho}_{23} \\ \tilde{\rho}_{31} \\ \tilde{\rho}_{32} \end{bmatrix} = - \begin{bmatrix} M_{19} \\ M_{29} \\ M_{39} \\ M_{49} \\ M_{59} \\ M_{69} \\ M_{79} \\ M_{89} \end{bmatrix}$$

(29.b)

To simplify the above expression, we define the following objects as before

$$B = \begin{bmatrix} \tilde{\rho}_{11} \\ \tilde{\rho}_{12} \\ \tilde{\rho}_{13} \\ \tilde{\rho}_{21} \\ \tilde{\rho}_{22} \\ \tilde{\rho}_{23} \\ \tilde{\rho}_{31} \\ \tilde{\rho}_{32} \end{bmatrix} \quad S = \begin{bmatrix} M_{19} \\ M_{29} \\ M_{39} \\ M_{49} \\ M_{59} \\ M_{69} \\ M_{79} \\ M_{89} \end{bmatrix} \quad W = \begin{bmatrix} (M_{11}-M_{19}) & M_{12} & M_{13} & M_{14} & (M_{15}-M_{19}) & M_{16} & M_{17} & M_{18} \\ (M_{21}-M_{29}) & M_{22} & M_{23} & M_{24} & (M_{25}-M_{29}) & M_{26} & M_{27} & M_{28} \\ (M_{31}-M_{39}) & M_{32} & M_{33} & M_{34} & (M_{35}-M_{39}) & M_{36} & M_{37} & M_{38} \\ (M_{41}-M_{49}) & M_{42} & M_{43} & M_{44} & (M_{45}-M_{49}) & M_{46} & M_{47} & M_{48} \\ (M_{51}-M_{59}) & M_{52} & M_{53} & M_{54} & (M_{55}-M_{59}) & M_{56} & M_{57} & M_{58} \\ (M_{61}-M_{69}) & M_{62} & M_{63} & M_{64} & (M_{65}-M_{69}) & M_{66} & M_{67} & M_{68} \\ (M_{71}-M_{79}) & M_{72} & M_{73} & M_{74} & (M_{75}-M_{79}) & M_{76} & M_{77} & M_{78} \\ (M_{81}-M_{89}) & M_{82} & M_{83} & M_{84} & (M_{85}-M_{89}) & M_{86} & M_{87} & M_{88} \end{bmatrix}$$

Substituting them into eqn. 29.b, we get

$$WB = -S \quad \text{or} \quad B = -W^{-1}S \qquad (30)$$

The Matlab program shown in Appendix B implements our algorithm for the three level system. Note that *this program is essentially the same as the program for the two-level case* with the following modifications: we have (a) defined additional parameters relevant to this system, (b) entered proper elements in the Hamiltonian, and (c) added appropriate source terms for the populations. As an example, we have shown in figure 4 a plot of the population of the excited state, produced using this code, displaying the well-known coherent population trapping dip.

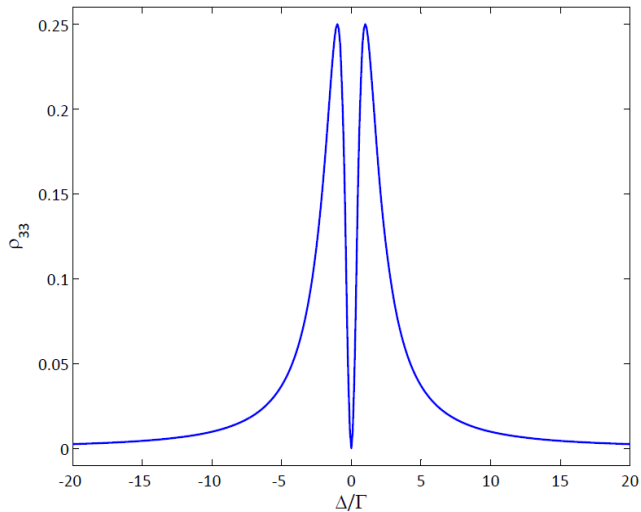

Figure 4. Population of the excited state for a three-level system calculated with this algorithm.

## 4. Applying the code to a system with an arbitrary number of energy levels

There are many examples in atomic and molecular physics where it is necessary to include a large number of energy levels. One example is an atomic clock employing coherent population trapping [12]. The basic process employs only three Zeeman sublevels. However, the other Zeeman sublevels have to be taken into account in order to describe the behavior of the clock accurately. Using alkai atoms for other applications such as atomic interferometry, magnetometry and Zeno-effect based switching also requires taking into account a large number of Zeeman sublevels [13-16 ]. Another example is the cooling of molecules using lasers. In this case, many rotational and vibrational levels have to be considered [17]. The code presented here can be applied readily to these problems, with the following modifications: (a) define additional parameters to characterize the problem, (b) develop the time independent Hamiltonian (possibly by inspection using the technique described earlier, if no closed-loop excitation is present), (c) add proper decay terms to the Hamiltonian, (d) add appropriate source terms for the populations & transverse decay terms, and (e) add plotting instructions for components of interest from the solution vector. Of course, if numerical techniques are to be employed for finding time-dependent solutions, the code can be truncated after the M matrix is determined, followed by application of eqn. 13 along with a proper choice of initial conditions.

## 5. Applying the code to a specific system with fifteen energy levels: an optically controlled waveplate

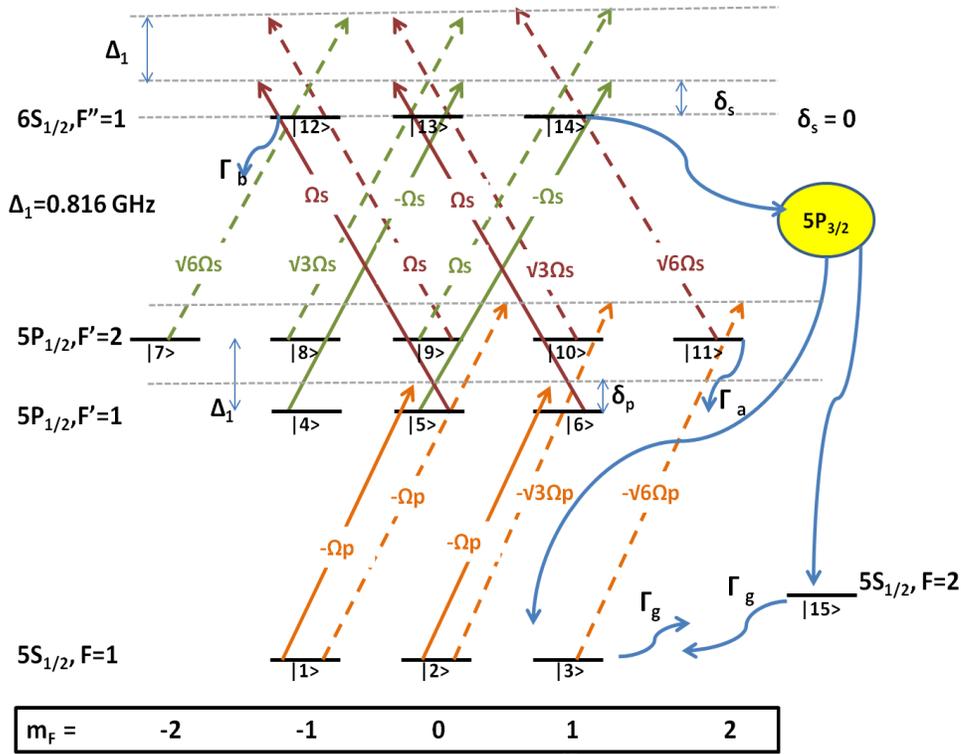

Figure 5. Fifteen level system for polarization rotation in $^{87}$Rb. |n>: Eigenstate of the system (n=1,2,…15); M: Zeeman sublevels (m=-2, -1, 0, 1, 2); The decay rates of $6S_{1/2}$ and $5P_{1/2}$ levels are $\Gamma_a$ and $\Gamma_b$ respectively. $\Gamma_g$: ground state dephasing rate. Rabi frequencies on the various legs are proportional to dipole strength matrix elements

As an explicit example of a system involving a non-trivial number of energy levels and optical transitions, we consider here a process where a ladder transition in $^{87}$Rb is used to affect the polarization of a probe beam (upper leg) by varying parameters for the control beam (lower leg). The excitation process is illustrated schematically in figure 5, for one particular configuration where the control beam is right circularly polarized, and the probe is linearly polarized. Because of the asymmetry introduced by the control, it is expected that the left circular component of the probe would experience a much larger phase shift, which in turn would induce an effective rotation of the probe polarization. Thus, the system can be viewed as an optically controlled waveplate for the probe. Here, we use the generalized algorithm to compute the response of this system. Of course, the response of the system under various experimental conditions would be quite different. The interactions of the pump (~795nm) and the probe (~1323nm) are modeled as follows. The pump is either left or right circularly polarized, and is tuned between the $5S_{1/2}$, F=1 -> $5P_{1/2}$, F'=1 and the $5S_{1/2}$, F=1 -> $5P_{1/2}$, F'=2 transitions, with a detuning of $\delta_p$, as illustrated in figure 5. The probe, linearly polarized, is tuned to the $5P_{1/2}$, F'=1 to $6S_{1/2}$, F"=1 transition, with a detuning of $\delta_s$. Due to Doppler broadening, it is important to consider the interaction of the $5P_{1/2}$, F'=2 level with both the pump and probe optical fields. For example, $\delta_p$ =814.5MHz corresponds to the situation where the pump is resonant with the $5S_{1/2}$, F=1 -> $5P_{1/2}$, F'=2 transition and $\delta_s$ =-814.5MHz corresponds to the situation where the probe is resonant with the $5P_{1/2}$, F'=2 to $6S_{1/2}$, F"=1 transition. In our model, we ignore the coherent coupling between $5S_{1/2}$, F=2 and the $5P_{1/2}$ manifold, because of the large frequency difference between $5S_{1/2}$, F=1 and $5S_{1/2}$, F=2 (~6.8GHz for $^{87}$Rb). However, we take into account the decay of atoms from the $5P_{1/2}$ manifold to the $5S_{1/2}$, F=2 state. Furthermore, we account for collisional relaxation (at a rate $\Gamma_g$) between $5S_{1/2}$, F=1 and 5S$_{1/2}$, F=2 manifolds, in order to model the behavior of atoms in a vapor cell. Finally, we also take into account the decay of atoms from 6S$_{1/2}$, F"=1 to the 5S$_{1/2}$ manifold via the 5P$_{3/2}$ manifold in an approximate manner.

The Rabi frequency of each transition is proportional to the corresponding dipole moment matrix elements. In Fig. 5, all the Rabi frequencies are expressed as a multiple of the Rabi frequency corresponding to the weakest transition [18]. For example, the dipole matrix elements of $\sigma^+$ transitions for the 5S$_{1/2}$-5P$_{1/2}$ excitation are tabulated in Table 1.

|  | $m_F = -1$ | $m_F = 0$ | $m_F = 1$ |
|---|---|---|---|
| $F' = 2$ | $-\sqrt{\frac{1}{12}}$ | $-\sqrt{\frac{1}{4}}$ | $-\sqrt{\frac{1}{2}}$ |
| $F' = 1$ | $-\sqrt{\frac{1}{12}}$ | $-\sqrt{\frac{1}{12}}$ |  |

Table 1: $^{87}$Rb D1 (5S$_{1/2}$ – 5P$_{1/2}$) Dipole Matrix Elements for $\sigma^+$ transition (F=1, $m_F \to F', m_F' = m_F + 1$)

Thus, if we set the coupling between |1> and |5> to be $\widetilde{H}_{1,5} = -\frac{\Omega_{P+}}{2}$, then the other coupling terms for the lower leg are as follows:

$\widetilde{H}_{1,9} = -\frac{\Omega_{P+}}{2}, \widetilde{H}_{2,6} = -\frac{\Omega_{P+}}{2}, \widetilde{H}_{2,10} = -\frac{\sqrt{3}\Omega_{P+}}{2}, \widetilde{H}_{3,11} = -\frac{\sqrt{6}\Omega_{P+}}{2}.$

The decay rates between any two Zeeman sub-levels are assumed to be proportional to the squares of the dipole moment matrix elements such that the sum of all the decay rates equals the net decay rate from that level. We assume all the Zeeman sub-levels in the 5P$_{1/2}$ and 6S$_{1/2}$ manifold decay at the same rate, $\Gamma_a$ and $\Gamma_b$ respectively. To illustrate how the decay terms are determined, consider, for example, state |5>, which denotes the Zeeman sublevel 5P$_{1/2}$, F'=1, m$_F$=0. The dipole matrix elements for all allowed transitions from this state to the various sublevels within the 5S$_{1/2}$ manifold are shown in Fig. 6. With the decay rate from |5> to the 5S$_{1/2}$ manifold being $\Gamma_a$, the decay rate from |5> to |1> (or |2>) is $\Gamma_a$/12. The decay from |5> to |15> (5S$_{1/2}$, F=2) is calculated by adding the squares of the matrix elements for all transitions between |5> and the Zeeman levels of |15>, and this turns out to be 5$\Gamma_a$/6.

Figure 6. Dipole Matrix Elements for all allowed transitions from the 5P$_{1/2}$,F'=1, m$_F$=0 sublevel to the various sublevels in the 5$2_{1/2}$ manifold.

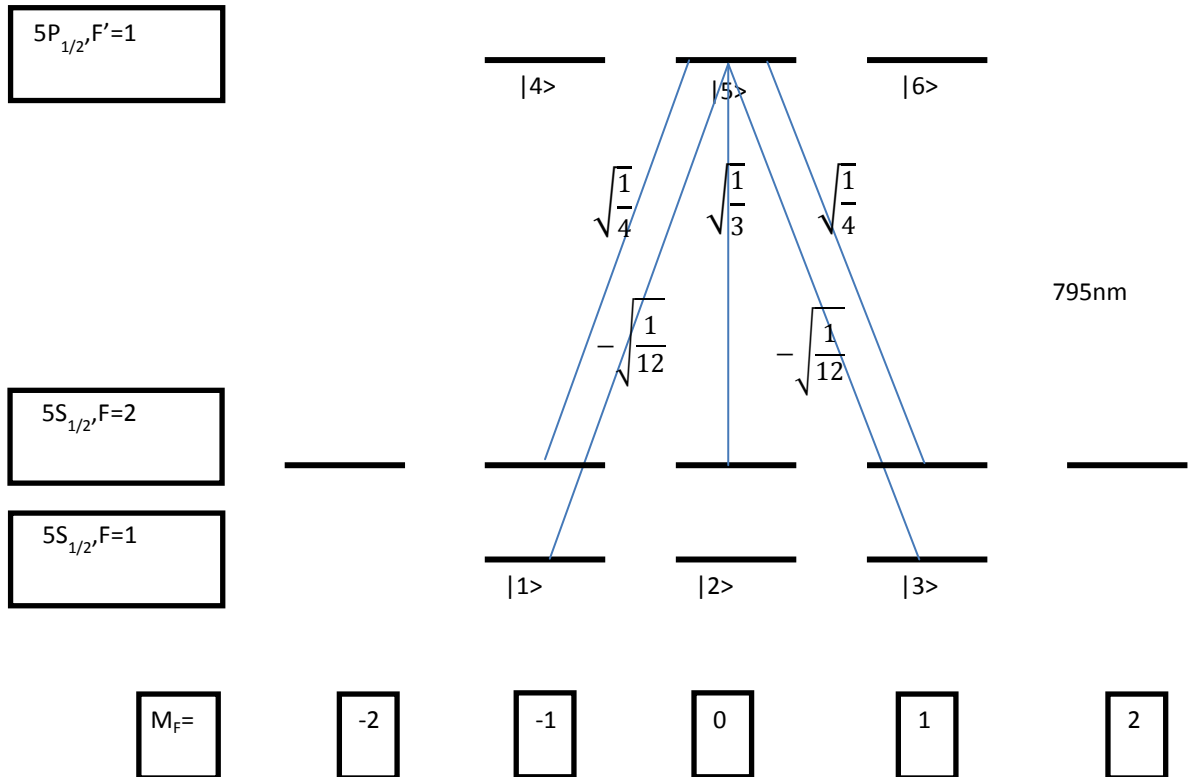

We have also taken into account the sourcing of atoms into the ground states from the $6S_{1/2}$ state via the $5P_{3/2}$ state. These additional source terms are modeled using an "effective decay rate" ($\Gamma_{bi}$) directly from the Zeeman sub-levels in the $6S_{1/2}$, F"=1 level to the $5S_{1/2}$ manifold. It is then assumed that all the Zeeman sub-levels at the $6S_{1/2}$, F"=1 level decays equally to the Zeeman sub-levels of F=1 and F=2 levels at this rate. In Fig 7, the branching ratios between the various hyperfine levels and the effective decay rates from the $6S_{1/2}$, F"=1 level to the $5S_{1/2}$ manifold are shown. For our initial computations, we used a rough estimate for $\Gamma_{bi}$. A more detailed calculation, taking into account the various branching ratios into and from all the hyperfine levels of the $5P_{3/2}$ state can be used to determine $\Gamma_{bi}$. However, we found that the results did not change significantly when $\Gamma_{bi}$ was changed slightly and hence using an approximate value is justified.

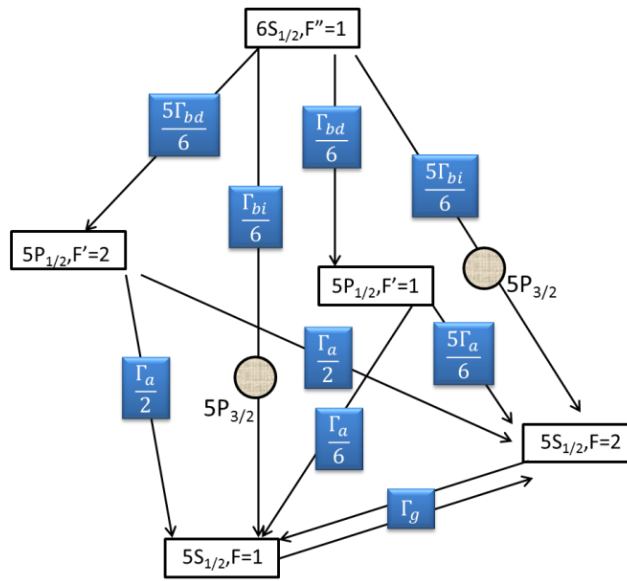

Fig. 7 Branching ratios between the hyperfine levels and the effective
decay rates from the F"=1 level to the 5S$_{1/2}$ manifold

The goal of the simulation of the process illustrated in figure 5 is to determine how the state of a linearly polarized probe beam (@1323nm) is affected by its passage through a vapor cell of length L and density n, in the presence of a circularly polarized pump beam (@795nm). Thus, before presenting the details of the atom laser interaction, we specify the terminology relevant for characterizing the probe beam, using the Jones vector formulation. We consider the direction of propagation as the z-axis, and the input probe to be linearly polarized in the x direction. Thus, the input probe can be described as:

$$\vec{J}_{\text{probe,input}} = \begin{bmatrix} 1 \\ 0 \end{bmatrix} = \frac{1}{2}\begin{bmatrix} 1 \\ i \end{bmatrix} + \frac{1}{2}\begin{bmatrix} 1 \\ -i \end{bmatrix} \quad (31)$$

The second part of eqn. (31) indicates that the linear polarization has been decomposed into a right circular polarization and a left circular polarization. The effect of propagation through the cell can now be modeled by expressing the output Jones vector as follows:

$$\vec{J}_{\text{probe,output}} = \frac{1}{2}\begin{bmatrix} 1 \\ i \end{bmatrix} e^{-\alpha_+ + j\Phi_+} + \frac{1}{2}\begin{bmatrix} 1 \\ -i \end{bmatrix} e^{-\alpha_- + j\Phi_-} \quad (32)$$

where $\alpha_+$ ($\alpha_-$) and $\Phi_+$ ($\Phi_-$) are the attenuation and phase shift experienced by the right (left) circular component, respectively.

In order to make the system behave as an ideal half waveplate, for example, the phase difference between the right and left polarization components ($|\phi_+ - \phi_-|$) should be equal to $\pi$, and the attenuation for each component should equal zero ($\alpha_+ = \alpha_- = 0$). In that case, the output expression can be simplified as:

$$\vec{J}_{\text{probe,output}} = \frac{1}{2}\begin{bmatrix} 1 \\ i \end{bmatrix} e^{j\phi_+} + \frac{1}{2}\begin{bmatrix} 1 \\ -i \end{bmatrix} e^{j\phi_-} = \frac{1}{2}e^{j\phi_-}\left(\begin{bmatrix} 1 \\ i \end{bmatrix} e^{j\pi} + \begin{bmatrix} 1 \\ -i \end{bmatrix}\right) = e^{j(\phi_- - \frac{\pi}{2})}\begin{bmatrix} 0 \\ 1 \end{bmatrix} \quad (33)$$

which is polarized linearly in the y-direction. In practice, the attenuation coefficients are non-vanishing. However, if they are equal to each other (i.e. $\alpha_+ = \alpha_-$), then they simply reduce the amplitude of the signal, without affecting the sense of polarization. Of course, the phase difference ($\Phi_+ = \Phi_-$) can have a wide range of values, corresponding to different output polarization states. In what follows, we solve the density matrix equation of motion for the 15-level system shown in figure 5, in order to determine the four quantities of interest: $\Phi_+, \Phi_-, \alpha_+, \alpha_-$.

The time-independent Hamiltonian after moving to a rotating basis and the RWA can be written down using the method we described in Section 2 & 3. Given the large number of levels, we use below a compact notation, rather than a matrix, to express the Hamiltonian. Specifically, $\widetilde{H}$ is given by (setting $\hbar=1$):

$\widetilde{H}_{1,1} = -i\frac{\Gamma_g}{2}, \widetilde{H}_{1,5} = -\frac{\Omega_P}{2}, \widetilde{H}_{1,9} = -\frac{\Omega_P}{2};$

$\widetilde{H}_{2,2} = -i\frac{\Gamma_g}{2}, \widetilde{H}_{2,6} = -\frac{\Omega_P}{2}, \widetilde{H}_{2,10} = -\frac{\sqrt{3}\Omega_P}{2};$

$\widetilde{H}_{3,3} = -i\frac{\Gamma_g}{2}, \widetilde{H}_{3,11} = -\frac{\sqrt{6}\Omega_P}{2};$

$\widetilde{H}_{4,4} = -\delta_P - i\frac{\Gamma_a}{2}, \widetilde{H}_{4,13} = -\frac{\Omega_S}{2};$

$\widetilde{H}_{5,1} = \widetilde{H}_{1,5}^*, \widetilde{H}_{5,5} = -\delta_P - i\frac{\Gamma_a}{2}, \widetilde{H}_{5,12} = \frac{\Omega_S}{2}, \widetilde{H}_{5,14} = -\frac{\Omega_S}{2};$

$\widetilde{H}_{6,2} = \widetilde{H}_{2,6}^*, \widetilde{H}_{6,6} = -\delta_P - i\frac{\Gamma_a}{2}, \widetilde{H}_{6,13} = \frac{\Omega_S}{2};$

$\widetilde{H}_{7,7} = \Delta - \delta_P - i\frac{\Gamma_a}{2}, \widetilde{H}_{7,12} = \frac{\sqrt{6}\Omega_S}{2};$

$\widetilde{H}_{8,8} = \Delta - \delta_P - i\frac{\Gamma_a}{2}, \widetilde{H}_{8,13} = \frac{\sqrt{3}\Omega_S}{2};$

$\widetilde{H}_{9,1} = \widetilde{H}_{1,9}{}^*, \widetilde{H}_{9,9} = \Delta - \delta_P - i\frac{\Gamma_a}{2}, \widetilde{H}_{9,12} = \frac{\Omega_S}{2}, \widetilde{H}_{9,14} = \frac{\Omega_S}{2};$

$\widetilde{H}_{10,2} = \widetilde{H}_{2,10}{}^*, \widetilde{H}_{10,10} = \Delta - \delta_P - i\frac{\Gamma_a}{2}, \widetilde{H}_{10,13} = \frac{\sqrt{3}\Omega_S}{2};$

$\widetilde{H}_{11,3} = \widetilde{H}_{3,11}{}^*, \widetilde{H}_{11,11} = \Delta - \delta_P - i\frac{\Gamma_a}{2}, \widetilde{H}_{11,14} = \frac{\sqrt{6}\Omega_S}{2};$

$\widetilde{H}_{12,5} = \widetilde{H}_{5,12}{}^*, \widetilde{H}_{12,7} = \widetilde{H}_{7,12}{}^*, \widetilde{H}_{12,9} = \widetilde{H}_{9,12}{}^*, \widetilde{H}_{12,12} = -\delta_S - \delta_P - i\frac{\Gamma_b}{2};$

$\widetilde{H}_{13,4} = \widetilde{H}_{4,13}{}^*, \widetilde{H}_{13,6} = \widetilde{H}_{6,13}{}^*, \widetilde{H}_{13,8} = \widetilde{H}_{8,13}{}^*, \widetilde{H}_{13,10} = \widetilde{H}_{10,13}{}^*, \widetilde{H}_{13,13} = -\delta_S - \delta_P - i\frac{\Gamma_b}{2};$

$\widetilde{H}_{14,5} = \widetilde{H}_{5,14}{}^*, \widetilde{H}_{14,9} = \widetilde{H}_{9,14}{}^*, \widetilde{H}_{14,11} = \widetilde{H}_{11,14}{}^*, \widetilde{H}_{14,14} = -\delta_S - \delta_P - i\frac{\Gamma_b}{2};$

$\widetilde{H}_{15,15} = -i\frac{\Gamma_g}{2}.$

All the other terms of $\widetilde{H}$ are equal to zero. We then add the population source terms to the Hamiltonian. We assume the decay rates from F"=1 to 5P$_{1/2}$ ($\Gamma_{bd}$) are equal to the effective decay rate from F"=1 to 5S$_{1/2}$ ($\Gamma_{bi}$). Thus, $\Gamma_{bd} = \alpha\Gamma_b$, $\Gamma_{bi} = (1-\alpha)\Gamma_b$ where $\alpha = 0.5$

$$\frac{d\rho_{11}}{dt} = (\rho_{44} + \rho_{55} + \rho_{99})\frac{\Gamma_a}{12} + \rho_{77}\frac{\Gamma_a}{2} + \rho_{88}\frac{\Gamma_a}{4} + (\rho_{12,12} + \rho_{13,13} + \rho_{14,14})\frac{\Gamma_{bi}}{18} + \rho_{15,15}\frac{\Gamma_g}{3}$$

$$\frac{d\rho_{22}}{dt} = (\rho_{44} + \rho_{66})\frac{\Gamma_a}{12} + \rho_{88}\frac{\Gamma_a}{4} + \rho_{99}\frac{\Gamma_a}{3} + \rho_{10,10}\frac{\Gamma_a}{4} + (\rho_{12,12} + \rho_{13,13} + \rho_{14,14})\frac{\Gamma_{bi}}{18} + \rho_{15,15}\frac{\Gamma_g}{3}$$

$$\frac{d\rho_{33}}{dt} = (\rho_{55} + \rho_{66} + \rho_{99})\frac{\Gamma_a}{12} + \rho_{10,10}\frac{\Gamma_a}{4} + \rho_{11,11}\frac{\Gamma_a}{2} + (\rho_{12,12} + \rho_{13,13} + \rho_{14,14})\frac{\Gamma_{bi}}{18} + \rho_{15,15}\frac{\Gamma_g}{3}$$

$$\frac{d\rho_{44}}{dt} = \rho_{12,12}\frac{\Gamma_{bd}}{12} + \rho_{13,13}\frac{\Gamma_{bd}}{12}$$

$$\frac{d\rho_{55}}{dt} = \rho_{12,12}\frac{\Gamma_{bd}}{12} + \rho_{14,14}\frac{\Gamma_{bd}}{12}$$

$$\frac{d\rho_{66}}{dt} = \rho_{13,13}\frac{\Gamma_{bd}}{12} + \rho_{14,14}\frac{\Gamma_{bd}}{12}$$

$$\frac{d\rho_{77}}{dt} = \rho_{12,12}\frac{\Gamma_{bd}}{2}$$

$$\frac{d\rho_{88}}{dt} = \rho_{12,12}\frac{\Gamma_{bd}}{4} + \rho_{13,13}\frac{\Gamma_{bd}}{4}$$

$$\frac{d\rho_{99}}{dt} = \rho_{12,12}\frac{\Gamma_{bd}}{12} + \rho_{13,13}\frac{\Gamma_{bd}}{3} + \rho_{14,14}\frac{\Gamma_{bd}}{12}$$

$$\frac{d\rho_{10,10}}{dt} = \rho_{13,13}\frac{\Gamma_{bd}}{4} + \rho_{14,14}\frac{\Gamma_{bd}}{4}$$

$$\frac{d\rho_{11,11}}{dt} = \rho_{14,14}\frac{\Gamma_{bd}}{2}$$

$$\frac{d\rho_{15,15}}{dt} = (\rho_{1,1} + \rho_{2,2} + \rho_{3,3})\Gamma_{gg} + (\rho_{44} + \rho_{55} + \rho_{66})\frac{5\Gamma_a}{6} + (\rho_{77} + \rho_{88} + \rho_{99} + \rho_{10,10} + \rho_{11,11})\frac{\Gamma_a}{2} + (\rho_{12,12} + \rho_{13,13} + \rho_{14,14})\frac{5\Gamma_{bi}}{6}$$

The attenuation and the additional phase shift introduced by the Rb medium (as compared to free space propagation) of the signal beam can be expressed as:

Phase:
$$\phi_+ = kL\frac{\beta_+}{2}Re(a_{13,4}\rho_{13,4} + a_{14,5}\rho_{14,5} + a_{12,7}\rho_{12,7} + a_{13,8}\rho_{13,8} + a_{14,9}\rho_{14,9})$$
$$\phi_- = kL\frac{\beta_-}{2}Re(a_{12,5}\rho_{12,5} + a_{13,6}\rho_{13,6} + a_{12,9}\rho_{12,9} + a_{13,10}\rho_{13,10} + a_{14,11}\rho_{14,11})$$

Attenuation:
$$\alpha_+ = e^{-kL\beta_+ Im(a_{13,4}\rho_{13,4}+a_{14,5}\rho_{14,5}+a_{12,7}\rho_{12,7}+a_{13,8}\rho_{13,8}+a_{14,9}\rho_{14,9})/2}$$
$$\alpha_- = e^{-kL\beta_- Im(a_{12,5}\rho_{12,5}+a_{13,6}\rho_{13,6}+a_{12,9}\rho_{12,9}+a_{13,10}\rho_{13,10}+a_{14,11}\rho_{14,11})/2}$$

And
$$\beta_\pm = b_{min}^2 \frac{3n_{atom}\Gamma\lambda^3}{4\pi^2\Omega_{min}},$$

where, $k$ is the wavevector of signal beam, which is at 1323nm, L is the length of the cell, which is set to be 15cm, $n_{atom}$ is the density of Rb atoms, which is set to be $10^{16}/m^3$, $\Omega_{min}$ is the Rabi frequency for the weakest probe transition (for example, the $|14\rangle$ - $|9\rangle$ transition in our model) and the various $a_{ij}$'s are the ratios of the Rabi frequency ($\Omega_{ij}$) of the $|i\rangle$-$|j\rangle$ transition to $\Omega_{min}$. For example, $a_{12,7} = \Omega_{12,7}/\Omega_{14,9} = \sqrt{6}$. $b_{min}^2$ is the fraction of the atoms (<1) that decay along the transition corresponding to $\Omega_{min}$, among all allowed decay channels from the decaying level. In our model, the amplitudes for all possible transitions from $|14\rangle$ are in the ratio $1:1:1:\sqrt{3}:\sqrt{6}$ and hence the fraction of atoms that decay along the different channels are in the ratio 1:1:1:3:6. Thus, $b_{min}^2 = 1/(1+1+1+3+6) = 1/12$.

Setting the pump frequency at a certain value ($\delta_p = \Delta$, which corresponds to the situation when the pump is resonant with the F=1 to F'=2 transition) and scanning the probe detuning ($\delta_s$), we can plot the various quantities of interest ($\Phi_+, \Phi_-, \alpha_+, \alpha_-$) as a function of $\delta_s$, as shown in Fig. 8. The relevant parameters used for this particular simulation are as follows. The decay rates $\Gamma_a$, $\Gamma_b$ and $\Gamma_g$ are $2\pi*5.75$ sec$^{-1}$, $2\pi*3.45$ sec$^{-1}$ and $2\pi*0.1$ sec$^{-1}$ respectively. We perform our calculations by setting $\Gamma_a$ to unity and rescaling all parameters in units of $\Gamma_a$. The separation $\Delta$, between F'=1 and F'=2 is $2\pi*814.5$ sec$^{-1}$ ($= 141.4\Gamma_a$) and the probe detuning ($\delta_s$) ranges from $-200\Gamma_a$ to $200\Gamma_a$. The Rabi frequencies have been chosen to be $\Omega_p = 5\Gamma_a$, and $\Omega_s = 0.1\Gamma_a$. Fig. 8(a) and 8(b) show the additional phase shifts produced by the Rb medium for the right and left circular polarization parts of the signal beam and Fig. 8(c) shows the difference between them. Fig. 8(d)-8(f) show the corresponding figures for attenuation. For example, at $\delta_s$=200, we have a differential attenuation of ~0 and a differential phase shift of about 30$^0$. Since the main purpose of this paper is to illustrate the application of the algorithm for obtaining the solution to the density matrix equations for a large quantum system, we refrain from exploring the parameter space in detail. Actual experimental results and conditions necessary to produce a differential phase shift of $\pi$ with virtually no differential attenuation (and thus allowing us to use the optically controlled waveplate for all-optical switching) are presented in a separate paper [19].

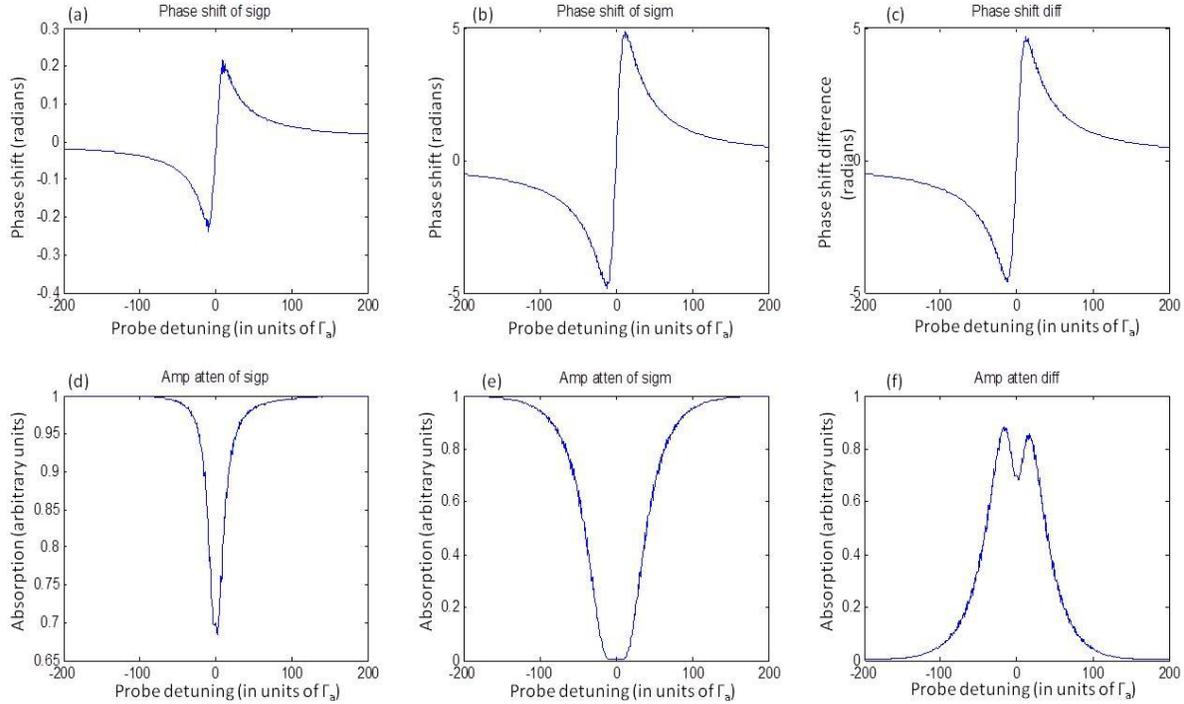
Figure 8. Simulation result of an optically controlled waveplate using 15 levels in $^{87}$Rb. See text for details.

## 5. Conclusion

We have presented a novel algorithm for efficiently finding the solution to the density matrix equations for an atomic system with arbitrary number of energy levels. For this purpose, the Liouville equation that describes the time evolution of the density matrix is formulated as a matrix-vector equation. We presented an algorithm that allows us to find the elements of the evolution matrix with ease for systems with arbitrarily large value of N. As examples, we then used the algorithm to find steady-state solutions for atomic systems consisting of two- and three-levels. We also described a comprehensive model (consisting of 15 levels) for an optically controlled waveplate using the $5S_{1/2}$-$5P_{1/2}$-$6S_{1/2}$ cascade system. Finally, we used the algorithm to obtain the steady state solution for the 15-level system. The algorithm and the Matlab codes presented here should prove very useful for the atomic and molecular physics community.

**Acknowledgements** This work was supported in part by AFOSR Grant numbers FA9550-10-01-0228 and FA9550-09-01-0652, NASA Grant number NNX09AU90A, NSF Grant number 0630388, and the DARPA ZOE program under Grant number W31P4Q-09-1-0014.

**Appendix A: Matlab Program for Solving the Two Level Problem**

```
omeg=5;            % express rabi freq, normalized to gamma
N=2;               % number of energy levels
R=401              % number of points to plot
                   % initialize and set dimensions for all matrices
delta=zeros(1,R);    %detuning array
M=zeros(N^2,N^2);    %M-matrix
rho=zeros(N,N);      %dens mat
Ham=zeros(N,N);      %Hamiltonian with decay
Q=zeros(N,N);        %matrix corresponding to derivative of the density matrix

W=zeros((N^2-1),(N^2-1));    %W-matrix
```

```matlab
S=zeros((N^2-1),1);         %S-vector
B=zeros((N^2-1),1);         %B-vector
A=zeros(N^2,R);             %A-vectors, for all detunings

for m=1:R      %start the overall-loop
 delta(1,m)=(m-(R+1)/2)/2; %define the detuning, normalized to gamma
 Ham=[0 omeg/2; omeg/2 (delta(1,m)+0.5i)*(-1)]; %elements of Hamiltonian

 for n=1:N^2     %start the outer-loop for finding elements of M;
     for p=1:N^2 %start inner-loop for finding elements of M;

                %M(n,p) equals Q(alpha,beta) with only rho(epsilon,
                %sigma)=1, and other elements of rho set to zero.

                %determining dummy coefficients alpha and beta
         remain=rem(n,N);
         if remain==0
             beta=N;
         else beta=remain;
         end
         alpha=(1+(n-beta)/N);

                %determining dummy coefficients epsilon and sigma
         remain=rem(p,N);
         if remain==0
             sigma=N;
         else sigma=remain;
         end
         epsilon=(1+(p-sigma)/N);

         rho=zeros(N,N);                %reset rho to all zeros
         rho(epsilon,sigma)=1;          %pick one element to be unity
         Q=(Ham*rho-rho*conj(Ham))*(0-1i); %find first part of Q matrix
         Q(1,1)=Q(1,1)+rho(2,2);        %add pop source term to Q
                                        %For an N-levl system, add additional
                                        %source terms as needed
         M(n,p)=Q(alpha,beta);
     end         %end the inner-loop for finding elements of M
 end             %end of the outer-loop for finding elements of M

         S=M(1:(N^2-1),N^2:N^2);       %find S-vector
         W=M(1:(N^2-1),1:(N^2-1));     %initialize W-matrix

         for d=1:(N-1)
             W(:,((d-1)*N+d))=W(:,((d-1)*N+d))-S; %update W by subtracting
                                                  %from selected columns
         end

         B=(W\S)*(-1);             %find B-vector: primary solution

         rhonn=1;                  %initialize pop of N-th state
                                   %determine pop of N-th state
         for f=1:(N-1)
             rhonn=rhonn-B(((f-1)*N+f), 1);
         end
```

```matlab
                                %determine the elements of the A vector
        A(1:(N^2-1),m)=B;
        A(N^2,m)=rhonn;

end             %end of over-all loop
plot(delta,real(A((N^2-0),:)))
```

## Appendix B: Matlab Program for Solving the Three Level Problem

```matlab
oma=1; omb=1;     % express omeg rabi freqs, in units of gamma
dels=0;           % common detuning set to zero
N=3;              % number of energy levels
R=401             % number of points to plot
                  %initialize and set dimensions for all matrices
del=zeros(1,R);    %diff detuning array
M=zeros(N^2,N^2);    %M-matrix
rho=zeros(N,N);      %density matrix
Ham=zeros(N,N);      %Hamiltonian with decay
Q=zeros(N,N);        %matrix representing derivative of density matrix
W=zeros((N^2-1),(N^2-1));    %W-matrix
S=zeros((N^2-1),1);          %S-vector
B=zeros((N^2-1),1);          %B-vector
A=zeros(N^2,R);              %A-vectors, for all detunings

for m=1:R      %start the overall-loop
 del(1,m)=(m-(R+1)/2)/10; %define the detuning
 Ham=[del(1,m)/2 0 oma/2; 0 del(1,m)*(-1)/2 omb/2; ...
     oma/2 omb/2 (dels+0.5i)*(-1)];

 for n=1:N^2      %start the outer-loop for finding elements of M;
     for p=1:N^2  %start inner-loop for finding elements of M;

                %finding alpha and beta
         remain=rem(n,N);
         if remain==0
             beta=N;
         else beta=remain;
         end
         alpha=(1+(n-beta)/N);

               %finding epsilon and sigma
         remain=rem(p,N);
         if remain==0
             sigma=N;
         else sigma=remain;
         end
         epsilon=(1+(p-sigma)/N);

         rho=zeros(N,N);               %reset rho to all zeros
         rho(epsilon,sigma)=1;         %pick one element to unity
         Q=(Ham*rho-rho*conj(Ham))*(0-1i); %find first part of Q matrix

         Q(1,1)=Q(1,1)+rho(3,3)/2;     %add pop source term to Q
         Q(2,2)=Q(2,2)+rho(3,3)/2;     %add pop source term to Q
                                       %Modify as needed for general
                                       %systems
         M(n,p)=Q(alpha,beta);
     end        %end the inner-loop for finding elements of M
 end            %end of the outer-loop for finding elements of M

         S=M(1:(N^2-1),N^2:N^2);       %find S-vector
         W=M(1:(N^2-1),1:(N^2-1));     %initialize W-matrix
```

```matlab
        for d=1:(N-1)
            W(:,((d-1)*N+d))=W(:,((d-1)*N+d))-S;  %update W by subtracting
                                                  %from selected columns
        end

        B=(W\S)*(-1);              %find B-vector: primary solution

        rhonn=1;                   %initialize pop of N-th state
                                   %determine pop of N-th state
        for f=1:(N-1)
            rhonn=rhonn-B(((f-1)*N+f), 1);
        end
                                   %determine elements of A vector
        A(1:(N^2-1),m)=B;
        A(N^2,m)=rhonn;

end                %end of over-all loop
plot(del,real( A ( (N^2-0),: ) ) )
```

## Appendix C: Algorithm Optimization

The crux of the algorithm is to obtain the M matrix in an automated fashion. The most obvious, but rather elaborate (O ($N^4$) operations) way to perform this task has been illustrated previously. However, several simplifications can be made to the algorithm so that the entire process can be accomplished using O ($N^2$) operations and also avoid some other redundant operations, thereby increasing the speed by a factor of ~$N^2$. To do this, we first observe that instead of evaluating the M matrix row-wise as was shown before, it is more beneficial to evaluate it column wise. Each column in the M matrix is simply obtained by successively setting each of the density matrix elements to 1, while setting all others to 0. Thus, the entire 1$^{st}$ column can be obtained be setting $\rho_{11}$=1 and all other $\rho_{ij}$=0, 2$^{nd}$ column with $\rho_{12}$=0 and all other $\rho_{ij}$=0 and so on. In general, by setting $\rho_{\varepsilon\sigma}$=1 and all other density matrix elements to 0, we obtain the (($\varepsilon$-1)*N+$\sigma$)$^{th}$ column of the M matrix where each of $\varepsilon$ and $\sigma$ vary from 1 to N.

Furthermore, it is to be noted that the computation H$\rho$-$\rho$H$^+$ involve multiplication of extremely sparse matrices, since only one of the elements of the $\rho$ matrix is 1 each time. It is evident that each column of the M matrix will simply be made up of certain columns of the Hamiltonian. Thus, the task is reduced to (a) figuring out the pattern of columns that are picked out from the Hamiltonian and (b) identify the locations in the M-matrix, where they would be filled. To illustrate this clearly, it is convenient to treat the calculation of the M-matrix as arising from two separate computations: H$\rho$ and $\rho$H$^+$. Let us consider a specific case when $\rho_{\varepsilon\sigma}$=1. The $\rho$H$^+$ computations would pick the $\sigma^{th}$ column of the Hamiltonian (with its elements conjugated) to be placed between rows ($\varepsilon$-1)*N+1 and $\varepsilon$*N of the (($\varepsilon$-1)*N+$\sigma$)$^{th}$ column of the M matrix. The H$\rho$ computations, on other hand, would pick the elements of the $\varepsilon^{th}$ column of the Hamiltonian (with the elements picking up an extra negative sign) and populate the following rows of the (($\varepsilon$-1)*N+$\sigma$)$^{th}$ column of the M matrix: $\sigma^{th}$ row, ($\sigma$+N)$^{th}$ row, ($\sigma$+2*N)$^{th}$ row and so on until the ($\sigma$+N*(N-1))$^{th}$ row. When, this process is repeated for each element of the density matrix, the M-matrix, barring the sourse terms would have been computed.

Finally, the addition of the source terms can also be simplified by choosing to modify the M-matrix only when one of the diagonal elements of the density matrix is set to 1, i.e $\rho_{\varepsilon\varepsilon}$=1, where $\varepsilon$=1 to N. Furthermore, instead of adding the source terms in-line, as was done previously, we can simply pre-define a "source-matrix" and simply pick off the elements of this matrix that would then be added to the appropriate entries in the M-matrix. For example, one way of defining such a "source matrix" would be to have the coefficients of the $\rho_{\varepsilon\varepsilon}$ in all the source equations (from $d\rho_{11}/dt$ to $d\rho_{NN}/dt$) along the $\varepsilon^{th}$ column of the source matrix. Now, all that needs to be done is to simply add the $\varepsilon^{th}$ column of the source matrix to the ($\varepsilon$-1)*N+ $\varepsilon^{th}$ column of the previously computed M matrix whenever $\rho_{\varepsilon\varepsilon}$=1. As an illustration of these optimization steps, we reproduce below a modified version of the code for a 3-level system, which should be contrasted with the unoptimized code for the same system presented in Appendix B.

```
oma=1; omb=1;     % express omeg rabi freqs, in units of gamma
dels=0;           % common detuning set to zero
N=3;              % number of energy levels
R=401             % number of points to plot
                  %initialize and set dimensions for all matrices
del=zeros(1,R);   %diff detuning array
M=zeros(N^2,N^2); %M-matrix
rho=zeros(N,N);   %density matrix
Ham=zeros(N,N);   %Hamiltonian with decay
```

```matlab
W=zeros((N^2-1),(N^2-1));    %W-matrix
S=zeros((N^2-1),1);          %S-vector
B=zeros((N^2-1),1);          %B-vector
A=zeros(N^2,R);              %A-vectors, for all detunings
Q_source=[0 0 1/2;
          0 0 1/2;
          0 0 0];

for m=1:R      %start the overall-loop
 del(1,m)=(m-(R+1)/2)/10; %define the detuning
 Ham=[del(1,m)/2 0 oma/2;
     0 del(1,m)*(-1)/2 omb/2; ...
     oma/2 omb/2 (dels+0.5i)*(-1)];

 col=0;  % index for column of M-matrix that will filled.
    index1=1:N;
    index2=1:N:N*(N-1)+1;
    index3=1:N+1:N^2;
    for n=1:N       %n keeps track of where in the M matrix the elements of
                    Ham have to be entered
        for p=1:N  %p picks the pth column from the Ham
           col=col+1;
           M(index1+(n-1)*N,col)=1i*conj(H(:,p));
           M(index2+p-1,col)=M(index2+p-1,col)-1i*(H(:,n));
           if n==p
               M(index3,col)=M(index3,col)+Q_source(:,n);
           end
        end            %end the inner-loop for finding elements of M
    end
    S=M(1:(N^2-1),N^2:N^2);      %find S-vector
    W=M(1:(N^2-1),1:(N^2-1));    %initialize W-matrix

    for d=1:(N-1)
        W(:,((d-1)*N+d))=W(:,((d-1)*N+d))-S; %update W by subtracting
                                             %from selected columns
    end

    B=(W\S)*(-1);           %find B-vector: primary solution!

    rhonn=1;                %initialize pop of N-th state
    for f=1:(N-1)
        rhonn=rhonn-B(((f-1)*N+f), 1);
    end

    A(1:(N^2-1),m)=B;
    A(N^2,m)=rhonn;
    M=zeros(N^2,N^2);
end             %end of over-all loop
plot(del,real( A ( (N^2-0),: ) ) )
```